\documentclass[aps,twocolumn,showpacs,superscriptaddress]{revtex4-1}
\usepackage{amsmath,amssymb}
\usepackage{graphics,graphicx}
\usepackage{dcolumn,bm}
\usepackage{psfrag}
\usepackage{subfloat}

\newcommand{\cu}
{\affiliation{Department of Physics, University of Calcutta,
92 Acharya Prafulla Chandra Road, Kolkata 700009, India.}}

\begin{document}

\title{Interplay of  interfacial noise and curvature driven dynamics in 
two dimensions}

\author{Parna Roy}
\cu
\author{Parongama Sen}
\cu

\begin{abstract}
We explore the effect of interplay of interfacial noise and curvature driven 
dynamics  in a binary spin system.  
An appropriate model is the   generalised two dimensional voter model  proposed  earlier (J. Phys. A: Math. Gen. {\bf 26},  2317 (1993)), where the flipping probability of a spin depends on the state of its neighbours and is given in terms of two parameters 
$x$ and $y$. $x = 0.5, y =1$ corresponds to the conventional voter model which is purely interfacial noise driven 
while $x = 1 $ and $y = 1$ corresponds to the Ising model, where coarsening is fully curvature driven. 
The coarsening phenomena for $0.5< x < 1$ keeping $y=1$ is studied in detail. 
The dynamical behaviour of the relevant quantities show characteristic differences from both $x=0.5$ and $1$.  
 The most remarkable result is the existence of two time scales for $x\ge x_c$ where $x_c \approx 0.7$.
 On the other hand,  we have studied the exit probability which shows Ising like behaviour with an universal 
exponent for any value of $x > 0.5$; the effect of $x$ appears in altering the value of the parameter occurring in the scaling function only. 
\end{abstract}

\pacs{89.75.Da, 64.60.De, 75.78.Fg}

\maketitle

Nonequilibrium phenomena associated with the zero temperature ordering process in classical Ising and Voter models \cite{liggett,krapbook,castellano,socio}
have been extensively studied in the recent past. Both models are two state models and the states can be represented by Ising spins. There is, 
however a basic difference.
The Ising model (IM) is defined using an energy function ($H = - J \sum \sigma_i \sigma_j$; where $\sigma = \pm 1$ and the sum is usually over nearest neighbours) and it has no intrinsic 
dynamics. However, starting from a configuration far from equilibrium, one can study the time dependent behaviour of the so called 
kinetic Ising model. At zero temperature, 
the time evolution  essentially corresponds to an energy minimising scheme \cite{Glauber} in the standard rules like single spin flip Glauber or Metropolis dynamics. The Voter model (VM) on the other hand has no such energy 
function associated - it is defined by the dynamical rule that an agent follows the state of a randomly chosen neighbour 
at each time step. The  kinetic Ising and Voter models  are known to be identical in one dimension 
while in higher dimensions the dynamical schemes are markedly different \cite{dornic}. 
While the coarsening is curvature driven in the Ising model, it is interfacial noise driven in the Voter model. 
This results in different behaviour of the relevant dynamical variables 
like density of active bonds $n(t)$, persistence probability $P(t)$ 
and time scales. Active bonds are those which connect neighbouring spins with opposite signs. 
In one dimension, for both the Ising
and Voter models, $n(t)$ shows power law decay as $t^{-\frac{1}{2}}$. This behaviour is  true   for the Ising model even in higher dimensions. But for the  Voter model, $n(t)$ 
asymptotically vanishes as $\frac{1}{\ln t}$ in two dimensions and for dimensions $d>2$, $n(t) \sim a-bt^{-d/2}$. The dynamics in VM is slower and the consensus time 
(by consensus we mean the all up and all down absorbing states of the system) 
typically behaves as $L^2 \log L$ in contrast to $L^2$ for the IM in two dimensions ($L$ is the system size). 
The persistence probability $P(t)$,  defined as the probability that a spin does not change sign till time $t$, shows 
algebraic decay as $t^{-\theta}$ with $\theta=0.375$ in one dimension \cite{Derrida1, Derrida2,Derrida3, satya} for the two models. 
For the the IM, $P(t)$ shows algebraic decay  even in higher dimensions; in two dimensions $\theta \approx 0.2$ \cite{bray,yurke,sire,picco}.
However, for the two dimensional VM, 
$ P(t)$  has the behaviour $\exp[-$constant$(\ln t)^{2}]$  \cite{ben}. 
The spin autocorrelation function  is another dynamical quantity which  again shows
different behaviour for the VM and the IM in two dimensions \cite{ben,auto1,auto2,auto3,auto4}.  

Another interesting feature of the $Z_2$ models (models with spin up and spin down symmetry)
with two absorbing states (which are either all spins up or down) is the exit probability $E(\rho)$. 
$E(\rho)$ is  defined as the probability that the  final  state has 
 all up spins starting 
with a density $\rho$ of up spins. The exit probability $E(\rho)$ is also different in the two models for $d>1$. It can be easily argued that 
in all dimensions,  $E(\rho)=\rho$ for the VM.  
With only nearest neighbour interaction, $E(\rho)$ is obviously linear for the IM also in one dimension. However, 
allowing further neighbour interactions and 
other parameters governing  the dynamics, a nonlinear behaviour can be 
observed even in one dimension for both Ising model and Voter model \cite{pr,lamred1,lamred2}. This implies that there is a scope for phenomena like ``minority spreading'' \cite{minor} always in one dimension. 
In the two dimensional Ising model  $E(\rho)$ shows non-linear behaviour \cite{Pm} which in the thermodynamic limit approaches a step function. It may be mentioned here that in the zero temperature 
ordering of Ising model, one encounters  the problem of frozen states \cite{redner,spirin}. Hence, the calculation of 
exit probability is made using only those configurations which reach the all up or all down states. 

Since  the interfacial noise and surface tension governed dynamics 
definitely  lead to highly different dynamical behaviour of several important quantities, it is worthwhile to study models in which both are present in a tunable manner. 
One such model had been proposed in \cite{oliveira}, namely the generalised voter model.
We have, therefore, considered this particular model to study the interplay of the interfacial noise and curvature driven dynamics.  

In the generalised voter model, the  dynamical rule has been parameterised so that one can recover a number of models for specific values of the parameters.
We have investigated the behaviour of the persistence probability, decay of active bonds, exit probability and time to reach consensus. 
These quantities are not related in general and hence the effect of changing the parameter values may be different for each of them.

Let us briefly review the generalised Voter model (GVM henceforth)  proposed in \cite{oliveira}. Here, at each site of the square lattice there is a spin variable $\sigma_i=\pm1$. 
The configuration evolves in time according to single spin flip stochastic dynamics. The spin flip probability $w_i(\sigma)$ for the $i$th spin is given by,
\begin{equation}
 w_i(\sigma)=\frac{1}{2}[1-\sigma_{i}f_i(\sigma)],
\end{equation}
where $f_i(\sigma)=f(\sum_{\delta}\sigma_{i+\delta})$, a function of the sum of the nearest neighbor spin variables. The model is defined 
 taking $f(0)=0$, 
$f(2)=-f(-2)=x$ and $f(4)=-f(-4)=y$, where $x$ and $y$ are restricted to the conditions $x\le 1$ and $y\le 1$. 
The original VM is recovered for $x=0.5$ and $y=1$ whereas the IM corresponds to  $x = 1, y=1$.  
Along the line $y=1$, there are two absorbing states: all spins up and 
all spins down for $x \geq 0.5$ (apart from possible frozen states). These states are however unstable for $x<0.5$. 
 As the limiting values $x=0.5$ and $x=1.0$ correspond to the two different models, 
 one can expect either a sharp transition or a crossover behaviour at an intermediate value of $x$.

We have studied the non-equilibrium behaviour and exit probability $E(\rho)$ of the GVM  keeping $y=1$ and varying $x$ using Monte Carlo simulations on $L \times L$ square lattices with $L \leq 80$. 
Periodic boundary conditions have been used and at least $2500$ different initial configurations 
have been simulated. Persistence probability, active bonds dynamics and consensus times are the quantities estimated. These quantities are unrelated and therefore it is useful to study all of these to check how each of them is effected 
by tuning of the parameter $x$.

\begin{figure}[!htbp]
\hspace*{-1.5cm}
\includegraphics[width=8.0cm,height=4.5cm,angle=0]{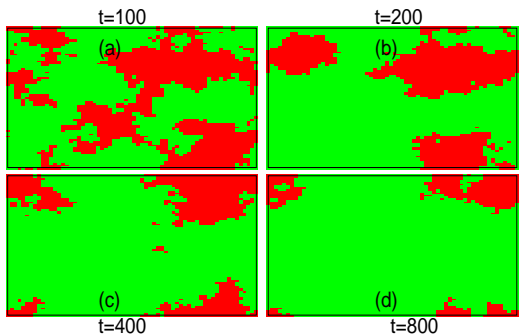}
\caption{Typical snapshots at different times for $x=0.6$ shows that
coarsening is curvature driven.}
\label{snp1}
\end{figure}

\begin{figure}[!htbp]
\hspace*{-1.5cm}
\includegraphics[width=8.0cm,height=4.5cm,angle=0]{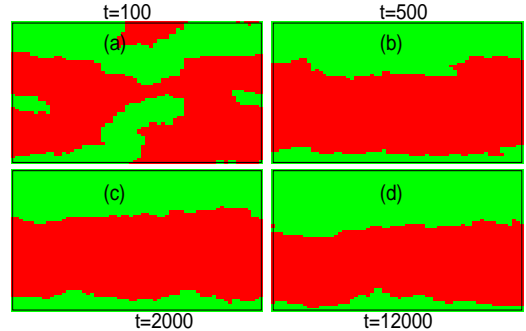}
\caption{Certain configurations show very slow 
relaxation as $x$ is increased.  Snapshots show such a configuration for different times for $x=0.9$.}
\label{snp2}
\end{figure}

Snapshots taken during the evolution help in understanding the process quite well. We find that in general for $x>0.5$ , the pictures looks very 
similar to the curvature driven case. However for $x$ close to $1$, certain configurations show the existence of nearly striped patterns  which do reach consensus 
but very slowly as the interfaces take long time to vanish. 
In Figs \ref{snp1} and \ref{snp2}, we have shown snapshots for two values of $x$; for $x = 0.6$, the curvature driven coarsening is seen to dominate while for $x = 0.9$, a case is shown where  coarsening has led 
to domains with nearly straight edges prevailing over large time durations.  Snapshots for other values of $x$ are given in the  \cite{SM}.

\begin{figure}[!htbp]
\hspace*{-1.5cm}
\includegraphics[width=8.0cm,height=4.5cm,angle=0]{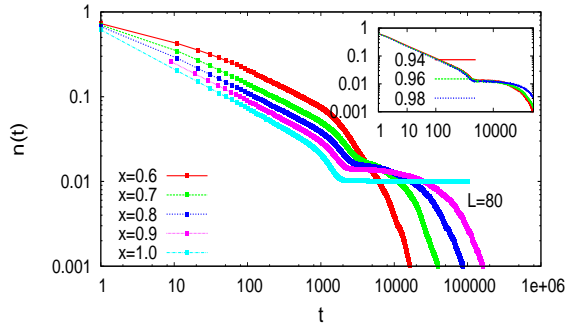}
\caption{Plot of density of active bonds with time $t$ for $L=80$ for $x=0.6, 0.7, 0.8, 0.9, 1$. 
Inset shows variation of active bonds for $x=0.94, 0.96, 0.98$.}
\label{dw}
\end{figure}

The variation of the density of active bonds $n(t)$ against time is plotted in Fig.~\ref{dw} for different values of $x$. 
As $x$ increases from $0.5$, we find that $n(t)$  goes to zero involving larger time scales. 
 However, the variation is faster than the $1/\log t$ behaviour known for 
the Voter model. 
 It is found that for any $0.5\le x<1$ the system always reaches the equilibrium 
ground state since $n(t)$ vanishes which implies the  freezing probability is zero. 
 Only at  $x=1$  a frozen state may be reached. 
As $x$ approaches unity (but not equal to it),  
the initial decay of $n(t)$ can be  fitted quite accurately by a power law, 
while there is a clear crossover to a much slower evolution  at later times 
 as shown in the inset of Fig. \ref{dw}.
This suggests that there are two different time regimes.
There is an  initial time scale up to which the behaviour is similar to the Ising model, i.e., $n(t)$ shows a  power law decay with exponent close to 1/2. 
Beyond this scale, a non algebraic slow decay is observed.   
However, we have checked that the behaviour in the later regime  is not like  $1/\log t$ 
as in the Voter model but may be even  slower than that. 
For general values of $x$, we conjecture that at initial times a power law behaviour occurs with some correction to scaling  as indicated  by the plots in Fig.~\ref{dw};
such corrections become weaker as   $x$ deviates from 0.5. 
The second regime with the slower decay exists only for  $x>x_c$, $x_c \approx 0.7$.
For exactly $x = 1$, the power law behaviour is exact before $n(t)$ saturates to a time independent non-zero value  due to 
the frozen stable states. 
%

\begin{figure}[!htbp]
\hspace*{-1.5cm}
\includegraphics[width=7.0cm,height=5.0cm,angle=0]{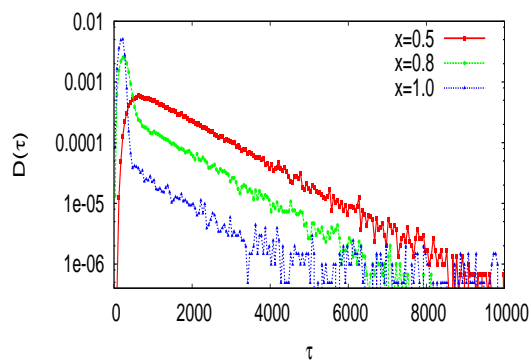}
\caption{ Plot of distribution of consensus time for $x=0.5, 0.8, 1$. }
\label{tau_dist}
\end{figure}

In order to gain more insight in the dynamical behaviour, 
we have estimated the   time $\tau$ required to reach the consensus state and its distribution $D(\tau)$. In a
detailed study made for $L=32$, we find that $D(\tau)$ changes its nature remarkably as $x$ is increased (Fig.~\ref{tau_dist}). For $x=0.5$, $D(\tau)$ shows a conventional behaviour; 
it increases for small $\tau$, has a broad peak and a long  exponential tail.  This behaviour continues till $x_c\approx 0.70$ beyond which we find that $D(\tau)$ differs considerably
for small and large values of $\tau$ (see Fig, 8 in SM). Apparently it is an overlap of a symmetric function of finite width peaked about a small value of $\tau$  
and a slow exponentially decaying  function extending to  large values of $\tau$ (a magnified figure is shown in Fig. 7 in the SM). Exactly at $x=1$,
the width of the symmetric function  is minimum  and the exponential part exists over 
a much shorter range. We conjecture that in the thermodynamic limit, the exponential part of $D(\tau)$ for $x=1$ will vanish altogether; this is  supported by the data for   $L = 64$  (shown  in \cite{SM}). 
As the  tail of the distribution for any $x$ may be  fit by an exponential function $\exp(-\omega\tau)$  one  can 
 define a  time scale $\tau_{eff} = 1/\omega$ for each $x$. 

\begin{figure}[!htbp]
\hspace*{-1.5cm}
\includegraphics[width=7.0cm,height=5.0cm,angle=0]{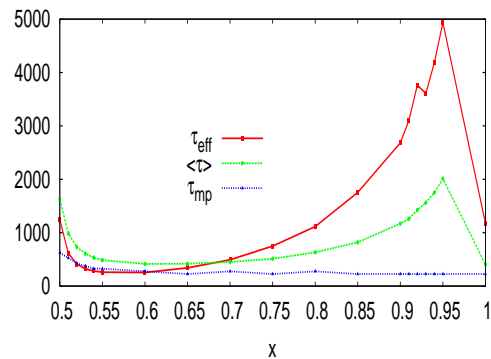}
\caption{ Plot of different time scales $\tau_{eff}$, $\langle\tau\rangle$ and $\tau_{mp}$ as a function of $x$.}
\label{tau_avg}
\end{figure}

 In Fig \ref{tau_avg}, we plot the three  time scales: $\tau_{eff}$, $\langle\tau\rangle$ and $\tau_{mp}$
where $\tau_{mp}$ denotes the most probable value.
In general,  the average values of $\tau$ are  different from the most probable values. 
 The difference becomes considerable for $x > x_c$  due to the long exponential tails. 
From Fig. \ref{tau_avg} we can conclude that $\tau_{mp}$ is very weakly dependent on   $x$; in fact for $x \geq 0.55$, it is almost a constant.
 On the other hand the two other timescales are strongly dependent on $x$;   $\tau_{eff}$ and $\langle\tau\rangle$ initially decrease and then increase  rapidly  with $x$.
Evidently, $\tau_{mp}$ denotes the time to reach consensus in absence of any intermediate metastable state and is  presumably constant  for any  
  $x> 0.5$.  On the other hand,  $\tau_{eff}$ corresponds  to the time scale associated with the 
configurations with nearly frozen intermediate states which increase in number as $x$ increases. $\langle \tau \rangle$ shows the 
increasing trend simply because it is an overall average. 

The existence of the two different time scales is clearly shown by 
the above result. 
 The non-monotonicity in $\langle \tau \rangle $ and  $\tau_{eff}$ are to be noted, both  time scales drop immediately as $x$ deviates from 0.5 and again at $x=1$. 
This  suggests  there are discontinuities at the two well known limiting points.

The average consensus times as a function of the different system sizes apparently  
 show a behaviour  faster than $L^{2}\log L$  for $x > 0.5$. $\langle\tau\rangle$ for other values of $L$ are shown in the \cite{SM}.  
The possibility of  discontinuities 
existing  at $x = 0.5$ and $x= 1$ is more strongly supported by the data as system size increases. 
A more conclusive statement about the scaling of the timescales
with the system sizes can only be made after a detailed study of 
the other time scales  which  is to be reported later.  

\begin{figure}[!htbp]
\hspace*{-1.5cm}
\includegraphics[width=7.5cm,height=5.0cm,angle=0]{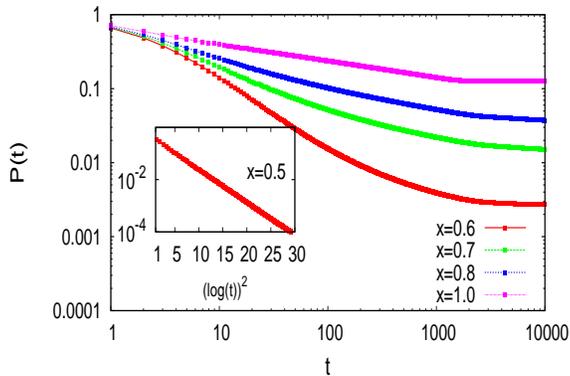}
\caption{Plot of persistence probability with time $t$ for $L=80$ for $x=0.6,0.7,0.8,1$. 
Inset shows variation of $P(t)$ as a function of $(\ln t)^2$ for $x=0.5$.}
\label{per}
\end{figure}

We next discuss some other results in context of the nonequilibrium phenomena. 
The persistence probability $P(t)$ as a function of time $t$ is plotted 
in Fig. \ref{per}. For $x=1$, $P(t)$ shows power law decay as $t^{-\theta}$ with 
$\theta\simeq0.2$ agreeing with the known result \cite{bray,yurke,sire,picco}. 
For $x=0.5$, persistence decays to a very small fraction ($\approx 0$) following the behaviour $\exp[-0.31(\ln t)^2]$. This behaviour 
is obtained by fitting the data and agrees very well with the form obtained numerically in \cite{ben}. 
For $0.5 < x < 1$, $P(t)$ also
approaches a non-zero saturation value, however there is no clear power law behaviour. 
The saturation value increases with increase in $x$ in a non-linear manner.

\begin{figure}[!htbp]
\hspace*{-1.5cm}
\includegraphics[width=7.5cm,height=5.0cm,angle=0]{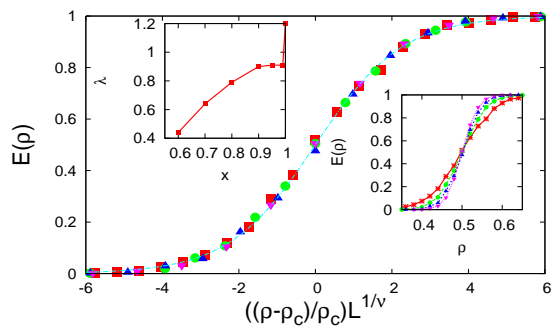}
\caption{Data collapse of $E(\rho)$ is plotted against $\frac{(\rho-\rho_c)}{\rho_c} L^{1/\nu}$ 
for system sizes $L=32,48,64,80$ for $x=0.6$. 
Bottom inset shows plot of unscaled data against $\rho$. Top inset shows the plot of $\lambda$ as a function of $x$.}
\label{ex_1}
\end{figure}

Lastly  we discuss the results for the exit probability (see Fig. \ref{ex_1}). 
The plot of  $E(\rho)$ as a function of $\rho$ shows that it is nonlinear except for $x=0.5$ having strong system size dependence. 
Different curves intersect at a single point $\rho=\rho_c \simeq 0.5$ ($\rho_c$ should be equal to 0.5 from symmetry argument). The curve becomes steeper as the system size is increased.
Finite size scaling analysis as in \cite{sbiswas} can be made using the scaling form 
\begin{equation}
 E(\rho,L)=f\left[\frac{(\rho-\rho_c)}{\rho_c} L^{1/\nu}\right],
 \label{scale}
\end{equation}
where $f(y) \to 0$ for $y<<0$ and equal to $1$ for $y>>0,$  so that the data for different system
sizes $L$ collapse when $E(\rho)$ is plotted against $\frac{(\rho-\rho_c)}{\rho_c} L^{1/\nu}$. The data collapse takes place with
$\nu = 1.3 ~\pm~0.01$ (the unscaled data is shown in the bottom inset) for all values of $x>0.5$. We conclude that like the Ising model, $E(\rho)$ becomes a step function in the thermodynamic limit. The 
scaling form given by eq. (\ref{scale}) is found to fit very well with a general form \cite{sbiswas}
\begin{equation}
 f(y)=\frac{1+\tanh(\lambda y)}{2},
 \label{ex1}
\end{equation}
where $\lambda$ depends on $x$. 
The dependence of $\lambda$ on $x$ is shown in the top left inset of Fig. \ref{ex_1}.  $\lambda$ which increases with $x$ quantifies the steepness of the $E(\rho)$ curve. 
 $\lambda$ increases   continuously as $x$ is increased from $0.5$,   consistent with the fact that the
$E(\rho)$ should  deviate from  the linear behaviour. However, exactly  at $x=1$, $\lambda$ increases abruptly to a comparatively larger value indicating a discontinuity.

Let us now discuss the results obtained in this Rapid Communication. First of all it appears that the curvature driven coarsening
governs the dynamics for any $x > 0.5$ at least in the initial stages. 
This is evident from  Fig. \ref{dw}, which shows that  at initial times the ordering becomes much faster  compared to the
Voter model 
as $x$ is increased from $0.5$ (see Fig. 5 in \cite{SM}).  
At the same time, the  interfacial noise driven coarsening present in the system, however small, is crucial for leading the system to 
consensus for larger values of $x$ when the dominant curvature driven process tends to generate nearly straight interfaces. 
It is only because of its presence the freezing probability is zero in the system for any $x \geq 0.5 $ (but not equal to unity). 
Thus it is interesting to note that while for smaller values of $x$  the model is closer to the voter model, the average consensus time
increases as $x$ approaches unity, the Ising limit.  This is  apparently counter intuitive as it is known that the evolution 
in the Voter model is slower compared to that in the Ising model in two dimensions. 
Actually the metastable states  increase in number as $x$ is increased (which is not surprising knowing the result for $x = 1$) enabling 
longer time scales for the system. However average consensus times are still less than that at $x=0.5$ up to a certain value of $x$. 
Results for different system sizes indicate that this value is very close to $x_c$ (shown in \cite{SM}) in the thermodynamic limit which is consistent with the other results.
 

The exit probability shows a nonlinear behaviour for any $x>0.5$ with a universal exponent $\nu~\approx1.3$ and a non-universal parameter $\lambda$ entering the 
scaling function. Since we omit the frozen states for $x=1$ and the time scales are irrelevant for this measure, 
it is not surprising that the behaviour is Ising like. However, the fact that  $\lambda$  
shows a discontinuity at $x=1$ again shows that  the $x=1$ point has  a distinctive feature  with respect to the exit probability as well.  

In conclusion, quite a few interesting results due to the interplay of the two types of dynamics are obtained in the genaralised Voter model. 
The main result is the existence of two time scales in the system for  $x> x_c$.One of them, $\tau_{mp}$ is  nearly independent of $x$ while $\tau_{eff}$, the other timescale is a highly nonlinear function of $x$.
 Both $x=0.5$ and $x=1$ with completely different kind of dynamical rules have unique features. At $x=0.5$, $P(t)$ and 
$E(\rho)$ behave differently compared to any other value of $x>0.5$, 
$\langle \tau \rangle$ is discontinuous. On the other hand, $x=1$ is also unique in the sense freezing occurs only at this point, $ \langle \tau \rangle$ and $\lambda$ show discontinuity here  
 while   well known power law behaviour in the relevant quantities exist.   
The intermediate region $0.5<x<1$
does not show any freezing phenomena. Although not algebraic, here $P(t)$ reaches saturation 
for $0.5<x<1$ in contrast to that at $x=0.5$. No sharp transition is observed for any value of $0.5<x<1$, but 
a crossover behaviour at $x=x_c\approx0.7$ is seen to exist.

Acknowledgement: PR acknowledges financial support from  UGC. PS acknowledges financial support from CSIR project.

\end{document}


\title{Interplay of interfacial noise and curvature driven coarsening:  Supplementary material}

\author{Parna Roy}
\cu
\author{Parongama Sen}
\cu

\maketitle

\section{Snapshots and the density of interfaces}

\begin{figure}[!htbp]
\hspace*{-1.5cm}
\includegraphics[width=8.0cm,height=4.5cm,angle=0]{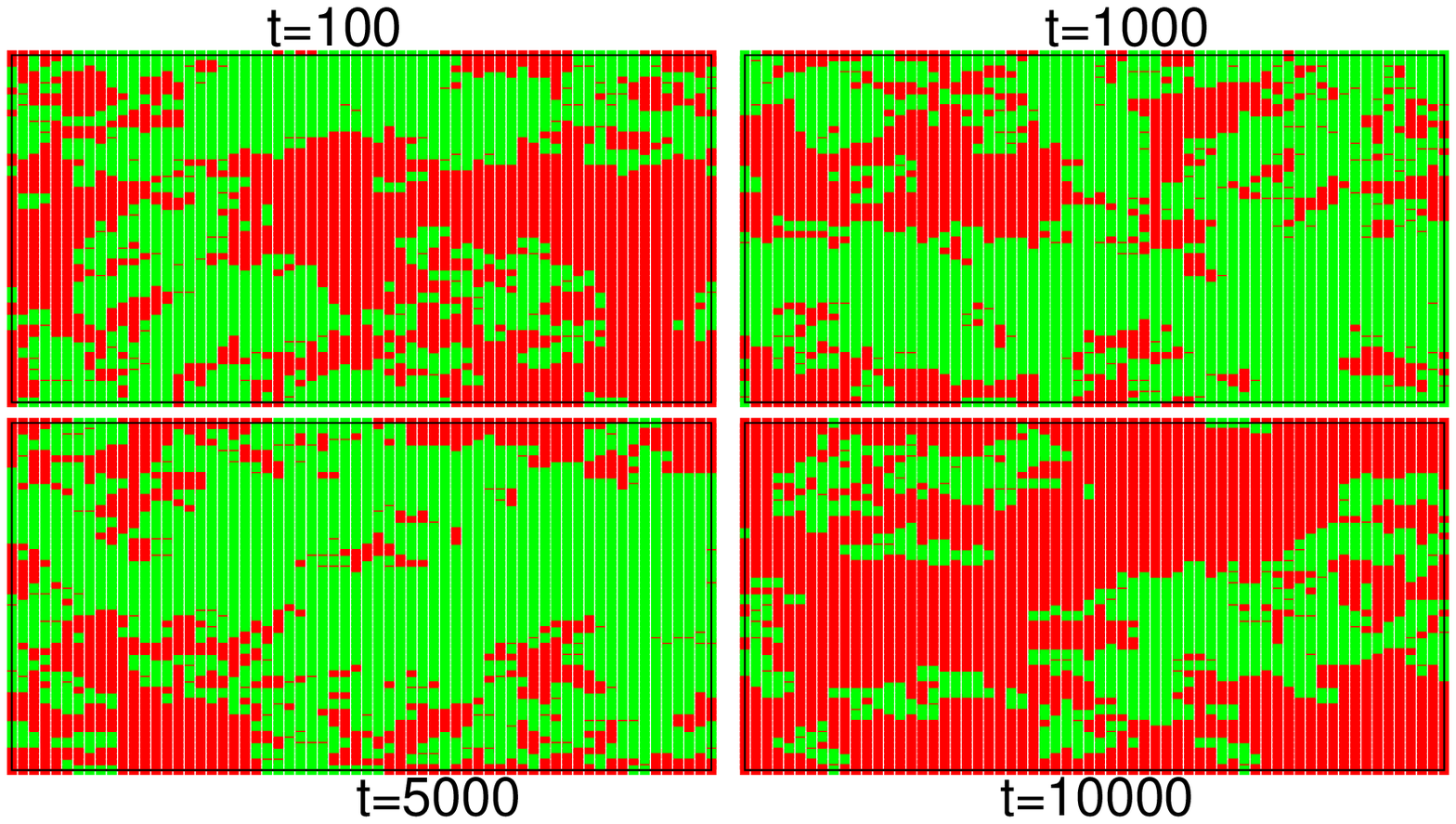}
\caption{Typical snapshots at different times for $x=0.5$ shows that
coarsening is interfacial noise driven.}
\label{snp0.5}
\end{figure}

The snapshots for different values of $x$ are plotted in the figures below for system size $64\times 64$. For $x=0.5$ (Fig. \ref{snp0.5}) 
we can see that there is no clear domain formation, only rough interfaces exist as is well known. 
Dynamical evolution   in this case is  interfacial noise driven; in a finite system, a random fluctuation of large size ultimately leads to a consensus state. 
For values of $x$ close to unity, we note that there are two possibilities. Either the system reaches a consensus state in a short time or it may take a much 
larger time as the system   evolves through  metastable states with minimum curvature. 

\begin{figure}[!htbp]
\hspace*{-1.5cm}
\includegraphics[width=8.0cm,height=4.5cm,angle=0]{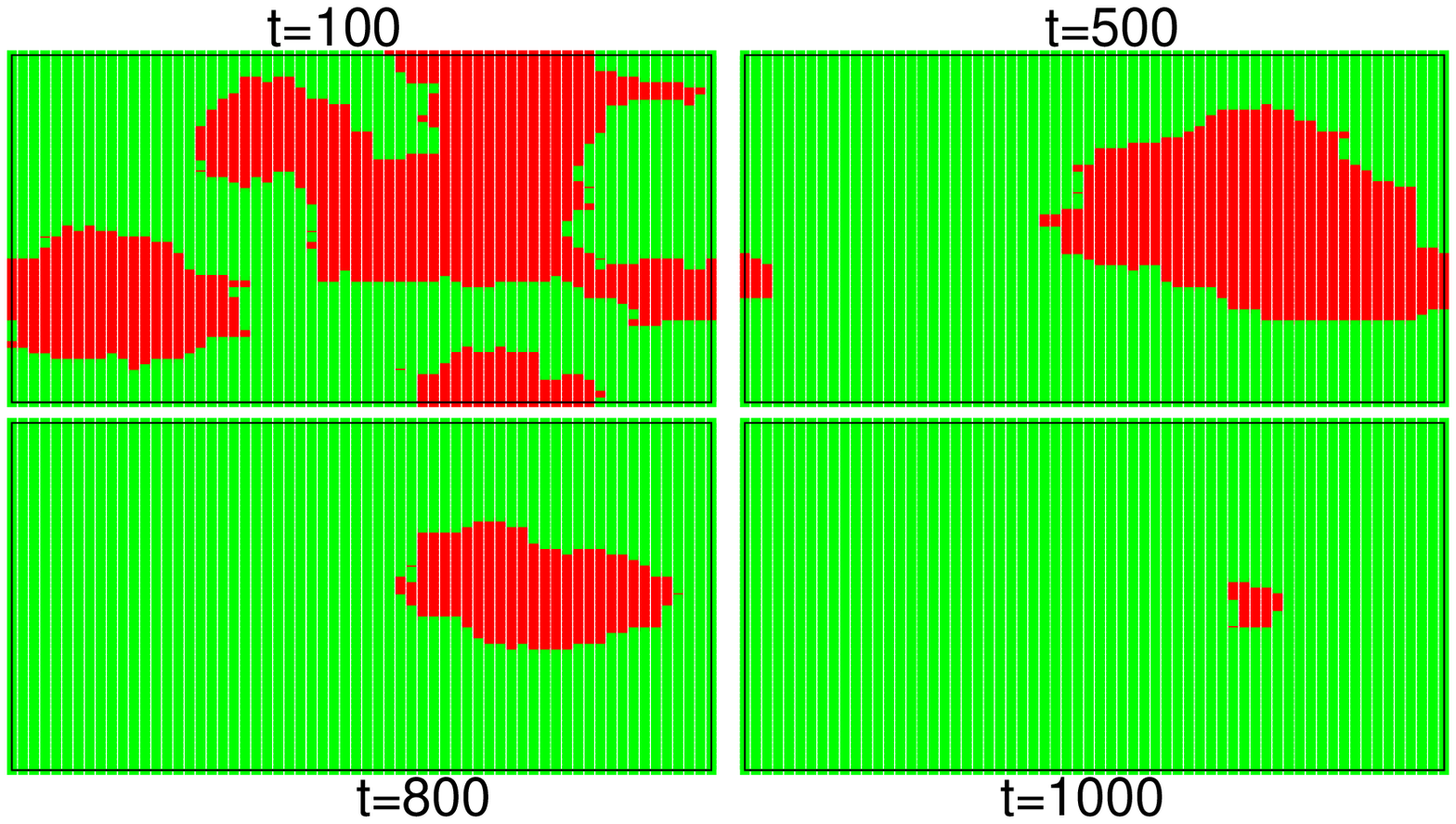}
\caption{Certain configurations show fast
relaxation.  Snapshots show such a configuration for different times for $x=0.9$.}
\label{snp0.9}
\end{figure}

\begin{figure}[!htbp]
\hspace*{-1.5cm}
\includegraphics[width=8.0cm,height=4.5cm,angle=0]{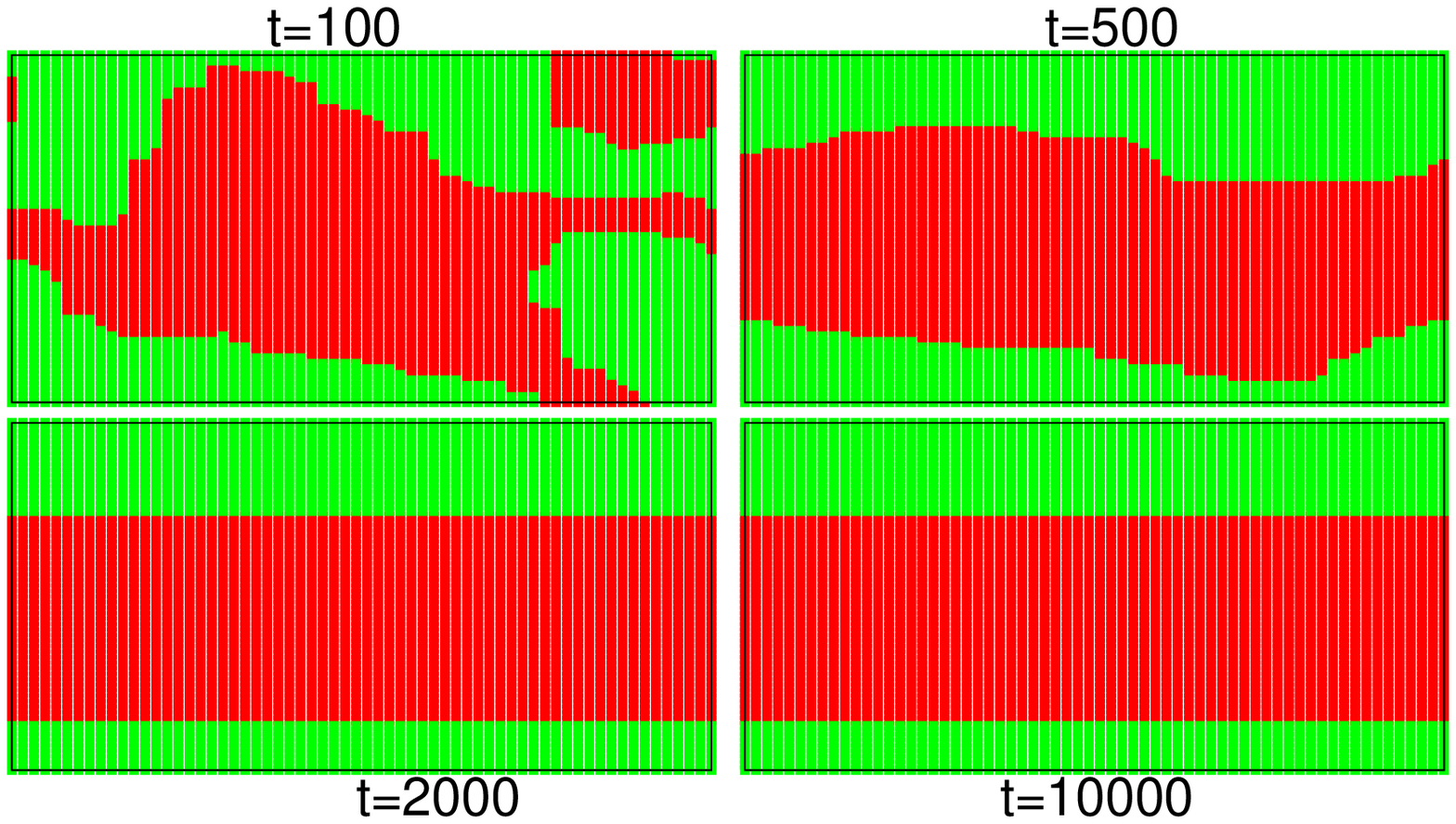}
\caption{Typical snapshots at different times for $x=1$ shows that
coarsening is curvature driven. This configuration reached striped frozen state.}
\label{snp1.0}
\end{figure}

\begin{figure}[!htbp]
\hspace*{-1.5cm}
\includegraphics[width=8.0cm,height=4.5cm,angle=0]{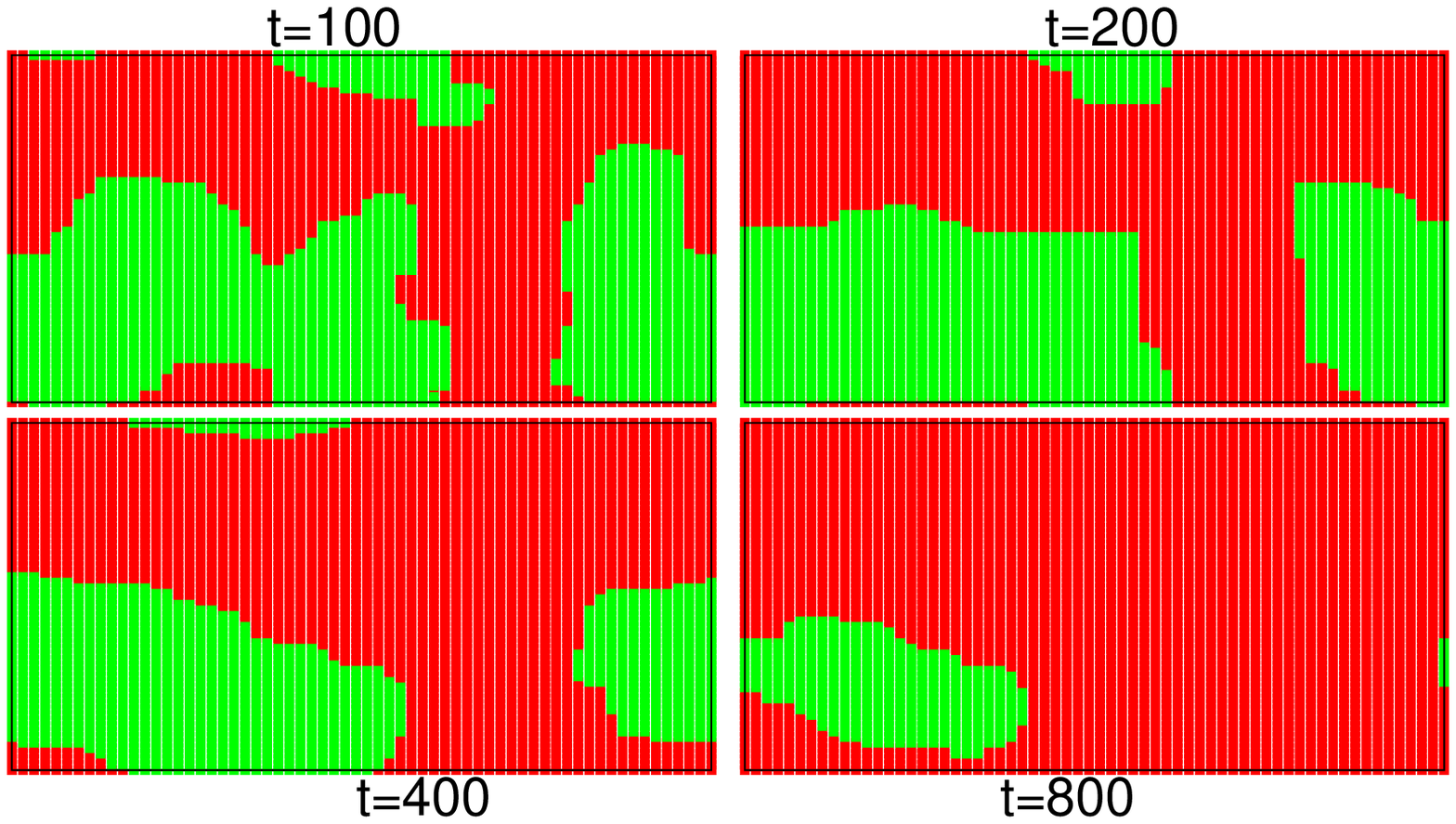}
\caption{Typical snapshots at different times for $x=1$ shows that
coarsening is curvature driven. This configuration reached the consensus state.}
\label{snp1.0_f}
\end{figure}

In the main text, we have shown the latter case for $x=0.9$. Here we show 
an example for the first case  (Fig. \ref{snp0.9}).
In this case no domains occur with nearly straight edges in the intermediate times. 
 For $x=1.0$ (Fig. \ref{snp1.0}) certain configurations reach frozen striped state which do not evolve further as the 
 coarsening is curvature driven.  In Figs \ref{snp1.0} and \ref{snp1.0_f}, 
two different cases of coarsening for $x=1$ are shown with or without freezing.  

\begin{figure}[!htbp]
\hspace*{-1.5cm}
\includegraphics[width=8.0cm,height=4.5cm,angle=0]{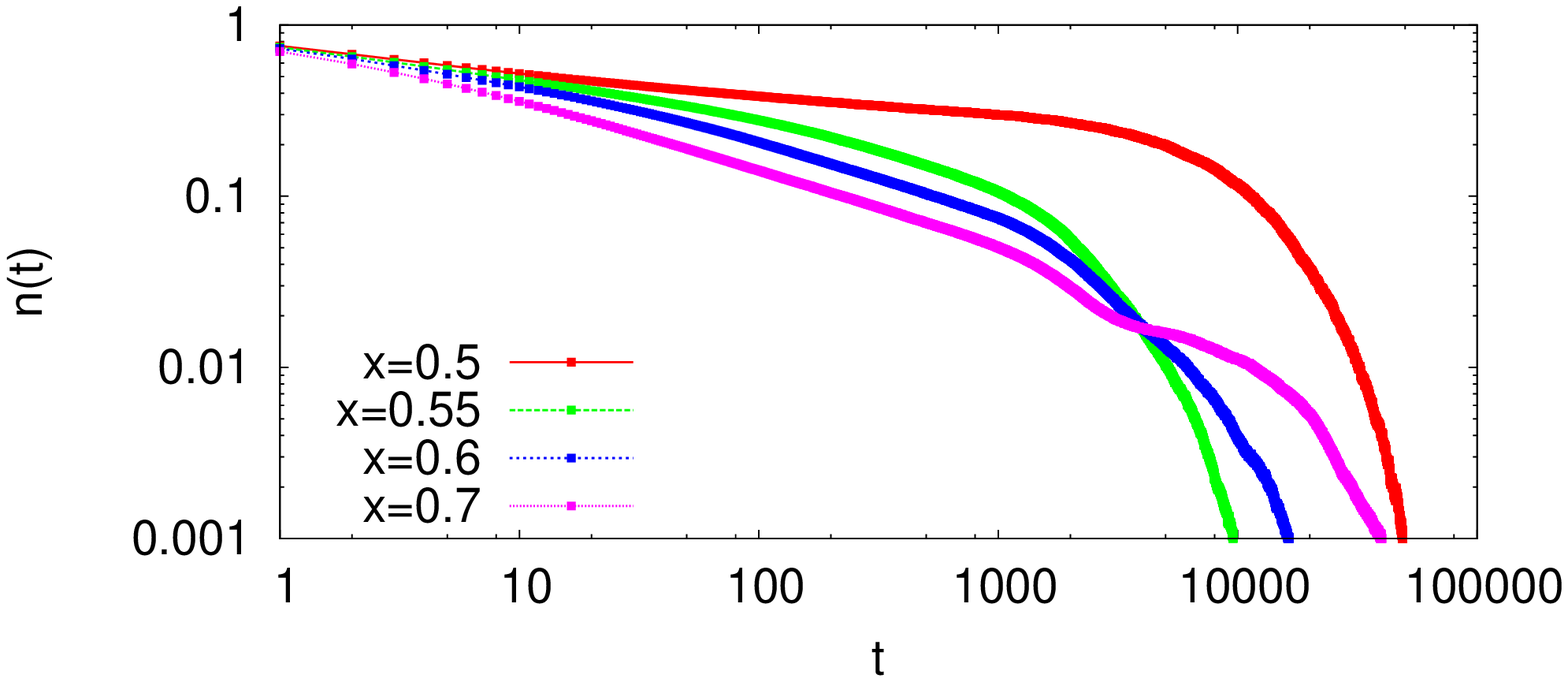}
\caption{Plot of density of interfaces $n(t)$ for $x=0.5,0.55,0.6,0.7$.}
\label{dw_supp}
\end{figure}

In Fig. \ref{dw_supp} we have plotted the density of interfaces $n(t)$ for values of $x = 0.5$ and a few other values.
In this  figure, it can be seen clearly that 
the coarsening becomes much faster compared to $x=0.5$ even 
as the deviation of $x$ from 0.5 is  small. 
It also shows that  a clear existence of a kink for $x = 0.7$ and not smaller values   
 supporting the conjecture that  a crossover behaviour occurs for $x \approx 0.7$.

\section{Consensus time}

\begin{figure}
\includegraphics[width=9.0cm,height=5.5cm,angle=0]{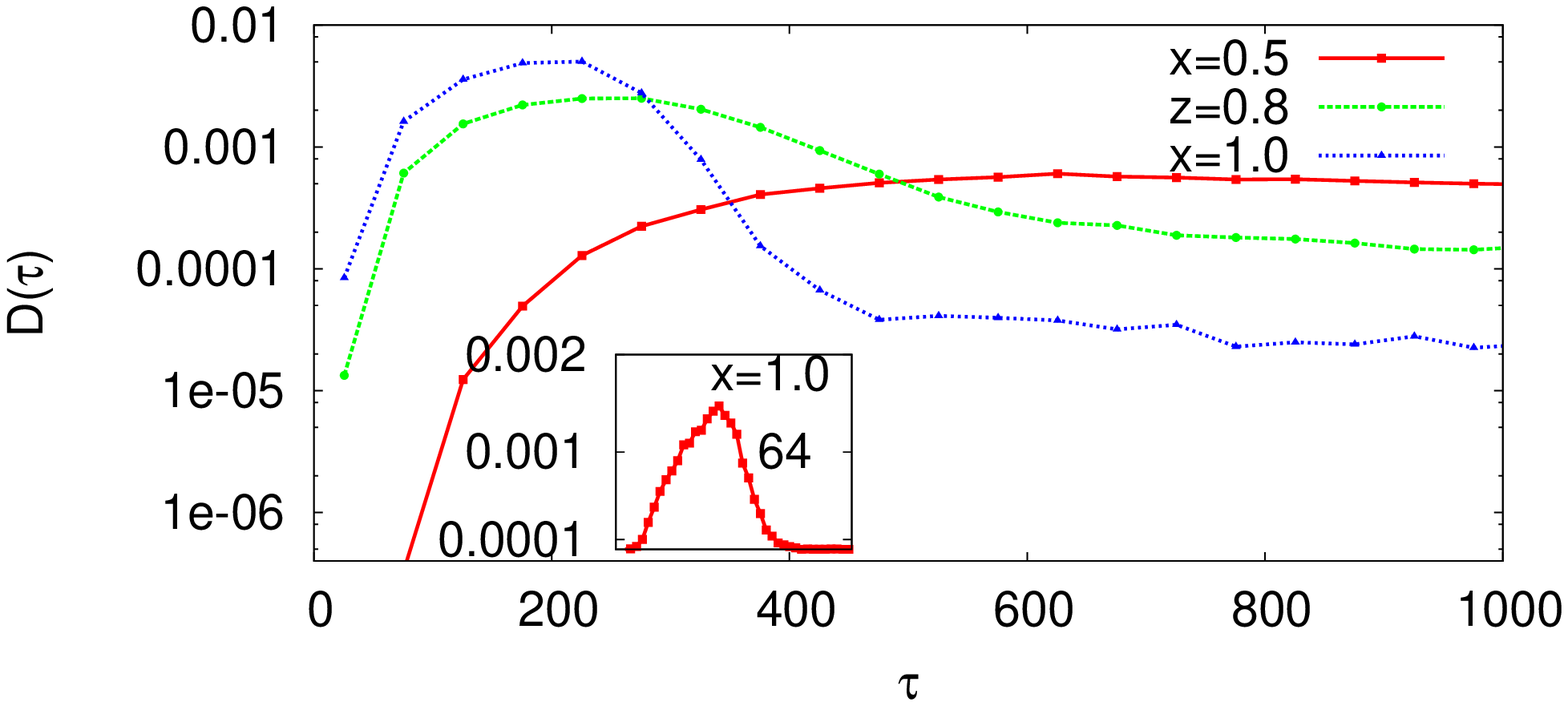}
\caption{Plot of $D(\tau)$ for initial time scale. Inset shows plot of the distribution of the consensus time $D(\tau)$ for $L=64$ for $x=1.0$.}
\label{dist_64}
\end{figure}

\begin{figure}
\includegraphics[width=9.0cm,height=5.5cm,angle=0]{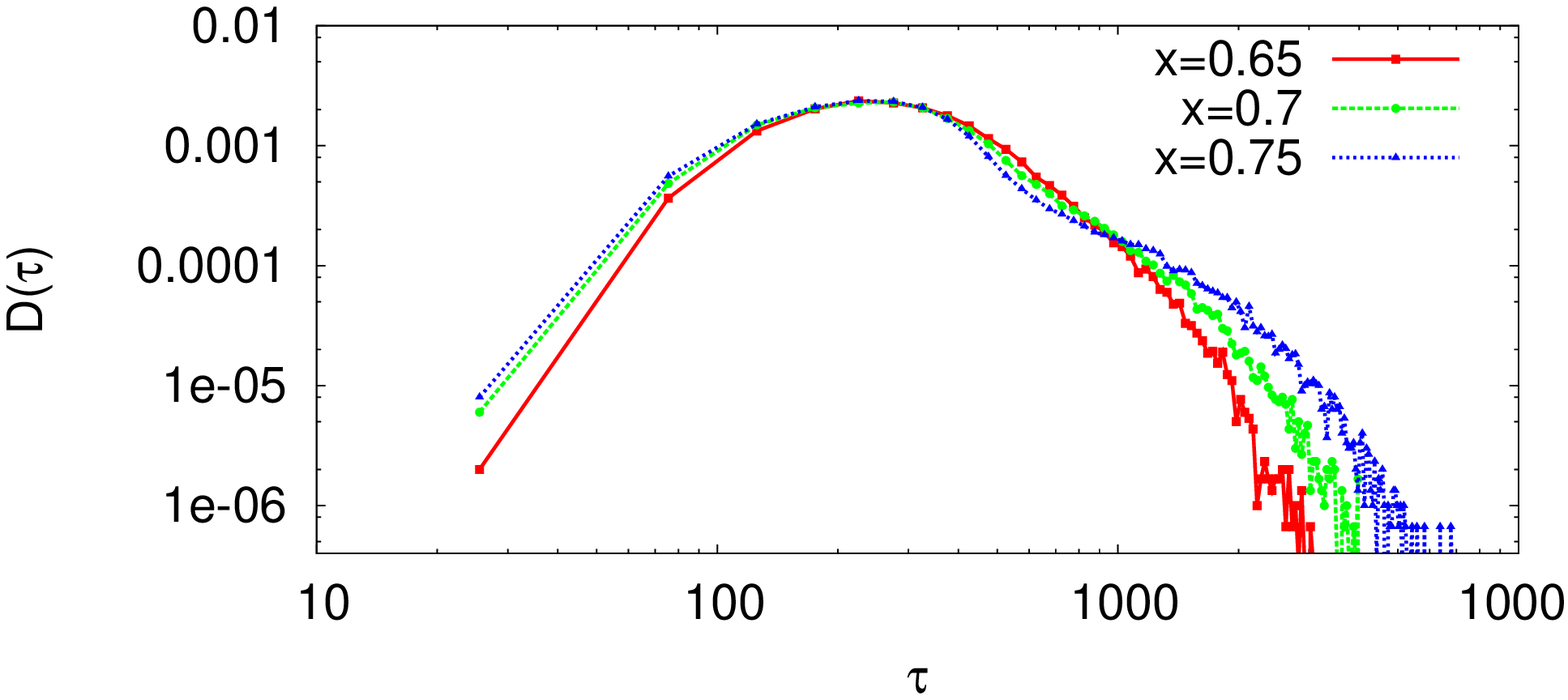}
\caption{Plot of $D(\tau)$ for $x=0.65, 0.7, 0.75$.}
\label{time1}
\end{figure}

In the main text, the consensus times distribution $D(\tau)$ for $L=32$ has been 
reported. 
In Fig. \ref{dist_64} we show  the distributions for $L=32$
magnifying the region $\tau \leq  1000$. 
small values of  $\tau$ only.  
 In the 
inset we have shown $D(\tau)$  for $x=1$ for $L=64$. From this figure we can argue that 
for larger system sizes, in the Ising limit $x=1.0$, the exponential decay at larger $\tau$ is not present, the distribution
only contains a sharply  peaked symmetric function of finite width. In the main text we have reported that the conventional behaviour 
of $D(\tau)$ continues till $x_c\approx0.7$ beyond which $D(\tau)$ changes its behaviour considerably. Fig. \ref{time1} supports the 
statement.

\begin{figure}
\includegraphics[width=9.0cm,height=5.5cm,angle=0]{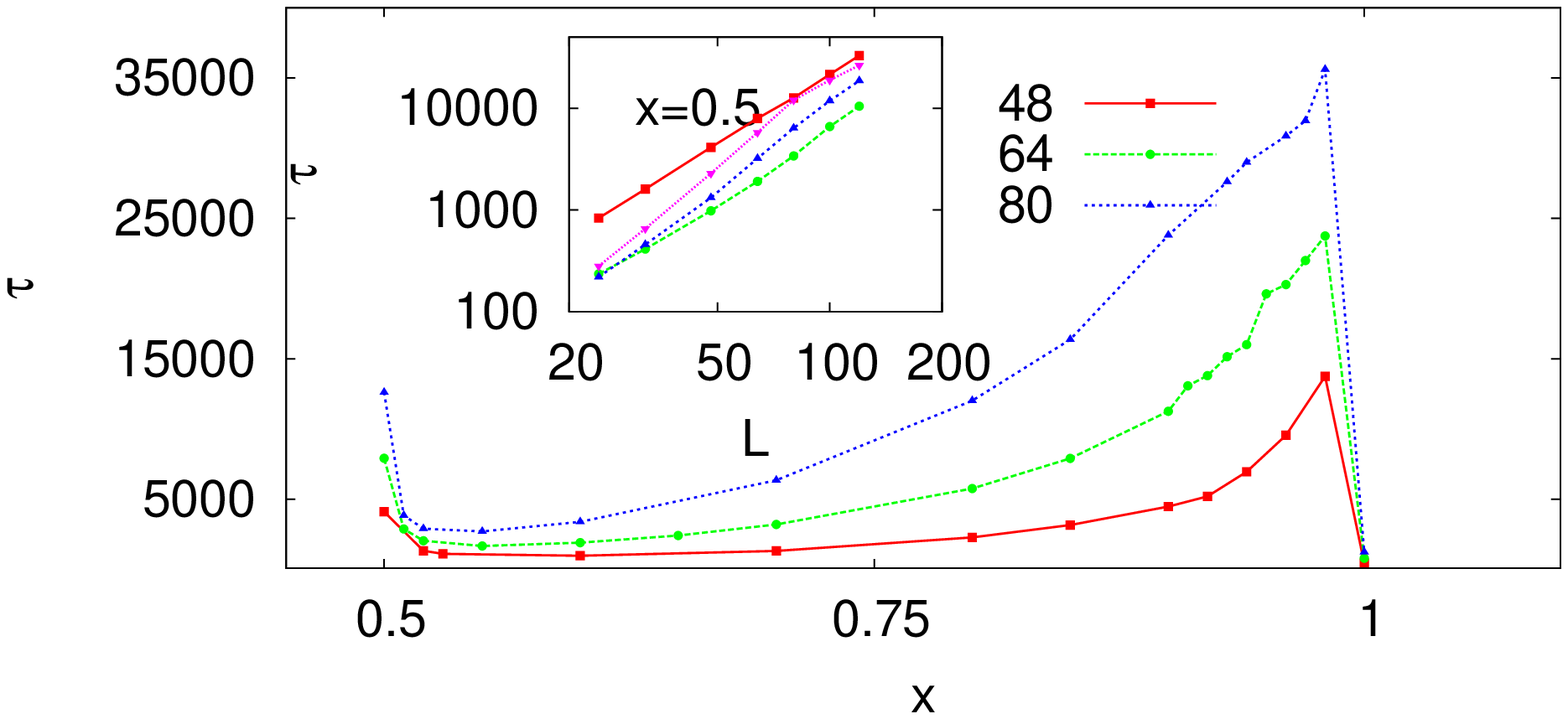}
\caption{Plot of consensus time $\tau$ as a function of $x$ for $L=48, 64, 80$. Inset shows variation of $\tau$ with system size for $x=0.5,0.6,0.7,0.8$.}
\label{time}
\end{figure}

For other values of $x$, 
only the average value of $\tau$  has been estimated so far for $L > 32$. 
In fig. \ref{time} we have plotted $\langle \tau \rangle$ as a function of $x$ for $L=48, 64, 80$. 
The results are qualitatively similar for $L=32$, and the results support the 
conjecture that discontinuities occur at $x=0.5$ and $1$. 
In the inset we have plotted $\langle \tau \rangle$ as a function of 
system size $L$
for different values of $x$.
Although for $x=0.5$ one gets the behaviour $\langle \tau\rangle \propto L^2 \log L$, it is difficult to conclude about the exact dependence for other values of $x$. For $x = 1$ the dependence is simply $L^2$ as is known. 


\begin{figure}
\includegraphics[width=8.0cm,height=4.5cm,angle=0]{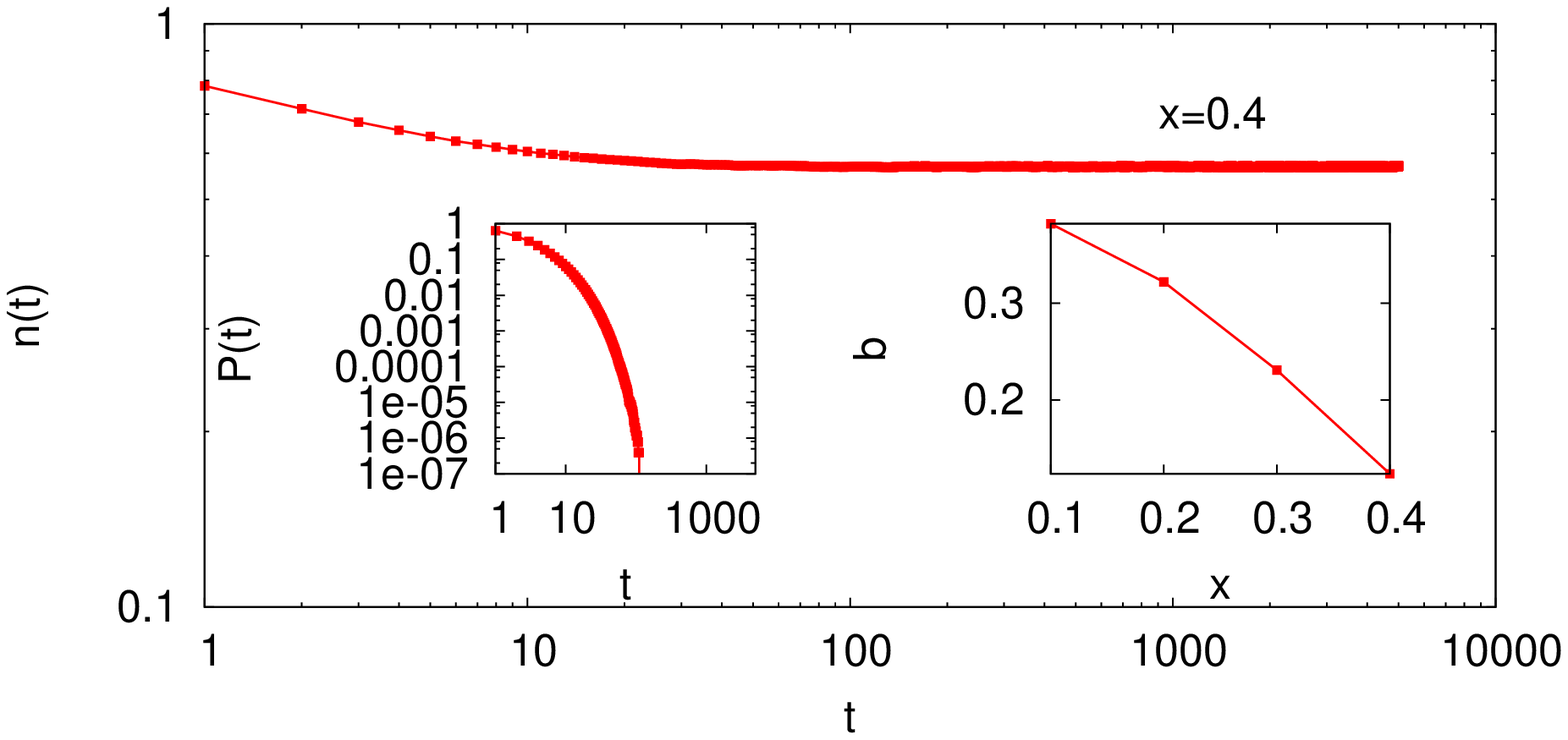}
\caption{Plot of $n(t)$ as a function of time for $L=32$ for $x=0.4$. Left inset shows variation of $P(t)$ with $t$ and right inset shows variation of the persistence 
exponent $b$ with $x$.}
\label{dw_per}
\end{figure}


\section{Results for $x<0.5$}

The focus of the paper has been on $x \geq  0.5$ for which absorbing states can be reached. However, the region $x < 0.5$ also yields some interesting results.  
As the state undergoes continuous evolution,  the persistence probability goes to zero
and density of active bonds remain finite. In Fig. \ref{dw_per} we have plotted density of active bonds $n(t)$ and in the inset we have 
plotted the persistence probability $P(t)$  as a function of time for $L=32$ for $x=0.4$.
The persistence probability has an exponential decay ($P(t)\sim\exp(-bt)$), i.e., it is faster than that in the voter model. 
The parameter $b$  has nonlinear dependence on $x$ (see inset of Fig. \ref{dw_per}).